\documentclass{aa}  

\usepackage{hyperref}
\usepackage[version=4]{mhchem}

\usepackage{upgreek}
\usepackage{soul}

\usepackage{graphicx}
\usepackage{txfonts}
\usepackage[dvipsnames]{xcolor}
\usepackage{placeins}

\newcommand{\chtwo}{CH$_2^+$}

\newlength{\dzero}
\newcommand{\minnumb}{\settowidth{\dzero}{$-$}\kern-\dzero$-$}
\newcommand{\spcn}[1]{\settowidth{\dzero}{0}\kern#1\dzero}

\begin{document} 
\definecolor{darkorange}{rgb}{1.0, 0.55, 0.0}
\definecolor{darkpastelgreen}{rgb}{0.01, 0.75, 0.24}

   \title{Searching for the elusive CH$_2^+$ with the James Webb Space Telescope}

   \subtitle{Another carbocation to constrain astrochemical networks}

   \author{M. Zannese \inst{\ref{iff},\ref{ias}}
          \and
          L. H. Coudert \inst{\ref{ismo}}
          \and 
          E. Dartois \inst{\ref{ismo}}
          \and P. Dell'Ova \inst{\ref{ias}} \and O. Roncero \inst{\ref{iff}} P. del Mazo-Sevillano \inst{\ref{salamanca}} \and U. Jacovella \inst{\ref{ismo}} \and B. Gans \inst{\ref{ismo}} \and J. R. Goicoechea \inst{\ref{iff}} \and D. Van De Putte \inst{\ref{uwo},\ref{uwo2},\ref{stsci}} \and C. Boersma \inst{\ref{nasa}} \and E. Habart \inst{\ref{ias}} \and E. Peeters \inst{\ref{uwo},\ref{uwo2},\ref{csc}} \and J. Cami \inst{\ref{uwo},\ref{uwo2}}\and R. Chown\inst{\ref{ohio}},   I. Schroetter\inst{\ref{irap}},  O. Kannavou\inst{\ref{ias}} }

   \institute{Instituto de F\'{\i}sica Fundamental
     (CSIC). Calle Serrano 121-123, 28006, Madrid, Spain \label{iff}\\\email{m.zannese@iff.csic.es}         
   \and
   Institut d'Astrophysique Spatiale, Universit\'e Paris-Saclay, CNRS,  Bâtiment 121, 91405 Orsay Cedex, France \label{ias}
    \and
    Institut des Sciences Moléculaires d’Orsay, UMR8214, CNRS, Université Paris-Saclay, 91405 Orsay, France \label{ismo}
\and 
Departamento de Qu{\'\i}mica F{\'\i}sica, Facultad de Ciencias Qu{\'\i}micas, Universidad de Salamanca
, 37008 Salamanca, Spain \label{salamanca}
              \and
              Department of Physics \& Astronomy, The University of Western Ontario, London ON N6A 3K7, Canada \label{uwo}\and Institute for Earth and Space Exploration, The University of Western Ontario, London ON N6A 3K7, Canada \label{uwo2}
              \and Space Telescope Science Institute, 3700 San Martin Drive, Baltimore, MD, 21218, USA \label{stsci}\and
        NASA Ames Research Center, MS 245-6, Moffett Field, CA 94035-1000, USA \label{nasa}
              \and Department of Astronomy, The Ohio State University, 140 West 18th Avenue, Columbus, OH 43210, USA \label{ohio} 
              \and Carl Sagan Center, SETI Institute, 339 Bernardo Avenue, Suite 200, Mountain View, CA 94043, USA \label{csc} 
              \and Institut de Plan\'etologie et d'Astrophysique de Grenoble (IPAG), Universit\'e Grenoble Alpes, CNRS, F-38000 Grenoble, France \label{irap}   
             }

   \date{Received April 15, 2026; accepted}

  \abstract
   {Carbocations are key species in 
   interstellar chemistry, providing entry points for building larger hydrocarbons. 
   CH$^+$ 
  , and more recently, CH$_3^+$, have been detected. Other carbocations await detection to provide a comprehensive view of the astrochemical network that is at work in the interstellar medium.}
   {We search for CH$_2^+$ in objects in which CH$_3^+$ was detected and evaluate the most favorable conditions for detecting the elusive \chtwo\ reactive cation.}
   {We calculated the CH$_2^+$ rotational and rovibrational transitions expected to contribute in the mid- to far-infrared, focusing on the lower-energy rovibrational levels. We then calculated CH$_2^+$ infrared emission spectra at different excitation temperatures and compared them to JWST spectra of the externally irradiated disk d203-506 in Orion, where $\rm CH^+$ and $\rm CH_3^+$ have already been detected. We used thermochemical models to predict the abundance and spatial morphology of CH$_2^+$ to better understand its nondetection.}
   {The comparison to JWST spectra allowed us to provide excitation-temperature-dependent upper limits to the  excited column density. These are several times lower than those detected for CH$^+$ and CH$_3^+$ in their excited states. Based on model calculations for photodissociation regions and assuming similar excitation temperatures, the upper limit derived from observations and CH$_2^+$ model spectrum is either slightly above or below the column density expected from models of photodissociation regions. 
   We provide a list of tabulated transitions to allow the community to search for this carbocation in future observations as CH$_2^+$ is key in providing
   observational constraints on astrochemical models.}
   {}


   \maketitle
   
   \nolinenumbers
%
%-------------------------------------------------------------------

\section{Introduction}

\begin{figure}
    \centering
    \includegraphics[width=\linewidth]{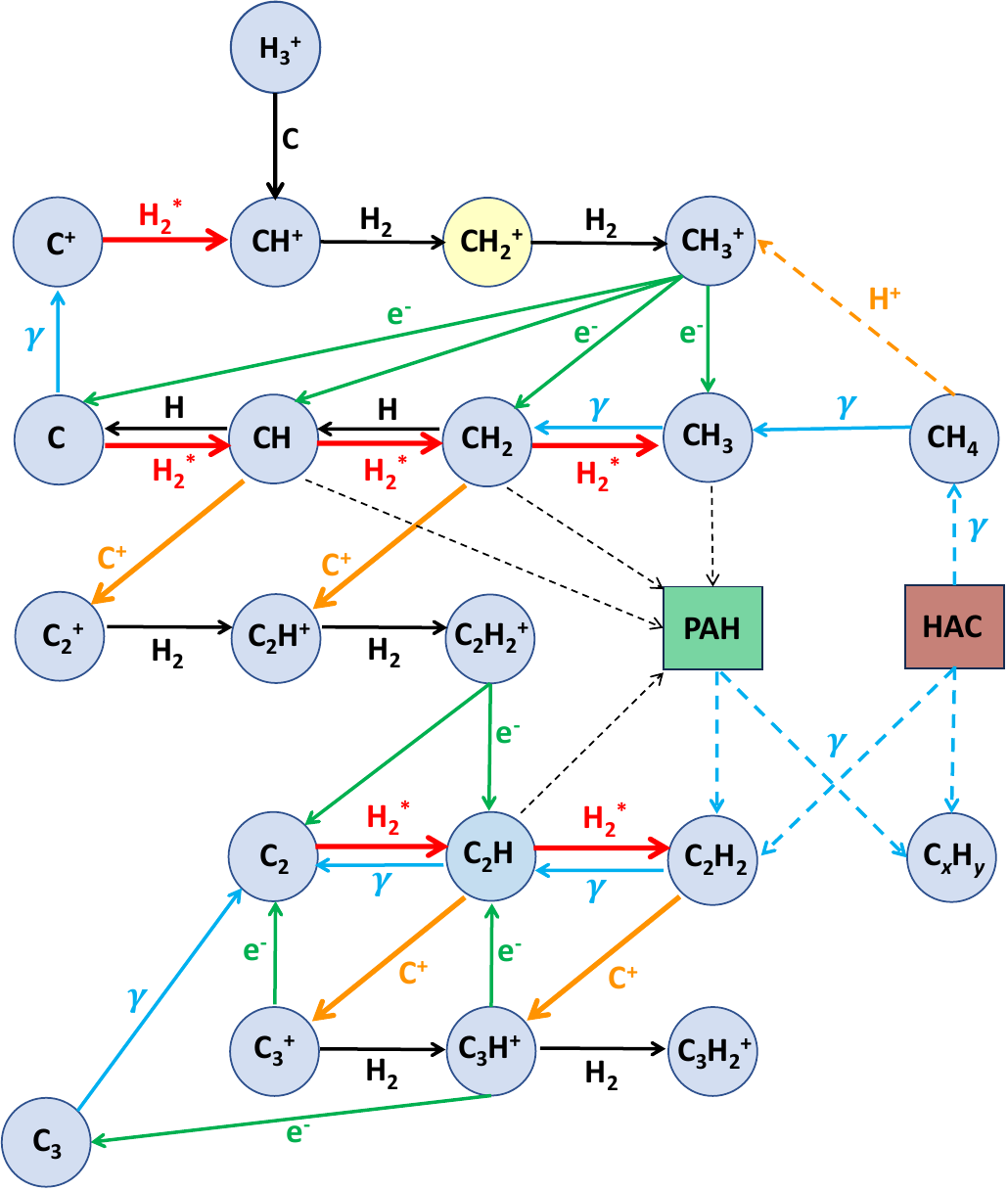}
    \caption{Carbon ion chemistry network. The red arrows indicate endoergic reactions, which proceed rapidly at high $T$ or in regions with significant FUV-pumped H$_2^*$. We also show possible reactions (dashed) involving Polyaromatic Hydrocarbon (PAH) and Hydrogenated Amorphous Carbon (HAC) grains, which may be relevant in certain conditions. The figure is adapted from \cite{Goicoechea_2025}.}
    \label{Figure_ion_chemistry}
\end{figure}

In dense interstellar clouds, the majority of small molecules are expected to be produced by gas-phase ion-neutral reactions (see Fig.~\ref{Figure_ion_chemistry}). The carbon ion chemistry is initiated by C$^+$, which is particularly abundant in ultraviolet (UV)-irradiated regions, or H$_3^+$, which is enhanced by cosmic rays as well as UV radiation \citep{Goicoechea_2025_h3+}. It is then followed by reactions with H$_2$ or C   
\citep{Smith1992,Indriolo2010,Herbst2021},

\begin{align}
    \ce{C+ ->[H_2] CH^+ ->[H_2] CH2+ ->[H_2] CH3+} \label{eq:C+}\\
    \ce{H3+ ->[C] CH^+ ->[H_2] CH2+ ->[H_2] CH3+} \label{eq:H3+}.
\end{align}

CH$_3^+$ then undergoes dissociative recombination, producing C, CH, or CH$_2$, which can then react with C$^+$ to produce larger molecules. In this context, CH$_2^+$ is an essential intermediate step in the carbocation chain leading to the formation of larger carbonaceous species. It thereby plays a pivotal role in astrochemical models as it is located in the molecular ladder between CH$^+$ and CH$_3^+$.

H$_3^+$ is now readily observed \citep{Geballe_1996,Geballe_2006} and offers a way to determine the degree of ionization in
diffuse clouds \citep[e.g.,][]{McCall_2002,Goto_2008,Indriolo_2012}.
The same holds for CH$^+$, which was observed a long time ago, first in absorption in the diffuse interstellar medium \citep{Douglas1941} and then in emission in various environments \citep[e.g.,][]{Cernicharo_1997,Naylor2010,Wesson_2010,Falgarone_2010,Bruderer_2010,Thi_2011,Spinoglio_2012,Rangwala_2014, Morris2016}. In contrast, $\rm CH_3^+$ has only recently been detected outside the Solar System, with the \textit{James Webb Space Telescope} (JWST) in the externally irradiated disk d203-506 near the Orion Bar \citep{Berne2023,Changala2023}. This was joined with the detection of the rovibrational emission of CH$^+$ in the disk and the Bar itself, which provided insight into the carbocation chain chemistry by unveiling correlations between H$_2$, CH$^+$, and CH$_3^+$, which confirms the consecutive hydrogen abstractions \citep{Zannese2025}. It also led to additional detections of CH$_3^+$ in UV-irradiated environments \citep{Henning2024,Bhatt_2025, Volz2026}. However, surprisingly, CH$_2^+$ remains undetected through its rovibrational emission expected in the mid-infrared  (JWST) and its rotational emission expected in the far-infrared (\textit{Infrared Space Observatory} and \textit{Herschel}/PACS).  
Upper limits on the abundance of CH$_2^+$ can thus provide additional constraints on astrochemical models (e.g., confirming the validity of the nondetection with models).

We present a calculation of the CH$_2^+$ infrared (IR) spectrum in the range spanned by JWST. We built a line-list, consisting of line positions (in cm$^{-1}$) and line intensities in LTE conditions, which allowed us to evaluate the observational constraints arising from the nondetection of CH$_2^+$ in objects in which CH$^+$ and CH$_3^+$ are detected.
In Sect.~\ref{sect:spectroscopy} we quantitatively describe the main features of the complex CH$_2^+$ spectroscopy. Sect.~\ref{sect:calculations} is devoted to the modeling of the CH$_2^+$ spectrum and to the building of a spectroscopic database for astrophysical purposes. In Sect.~\ref{sect:comp_JWST} we constrain the nondetection of CH$_2^+$ emission in d203-506, a reference target in which CH$_3^+$ is well detected, and we derive an upper limit of its excited column density. In Sect. \ref{sect:PDR_models} we present a fiducial Photodissociation Regions (PDR) model of d203-506 to analyze the predicted abundance and morphology of CH$_2^+$. We then discuss the results and conclude with future prospects for CH$_2^+$ detection.

%
%________________________________________________________
\begin{figure}
\begin{center}
\includegraphics[width=\columnwidth,angle=0]{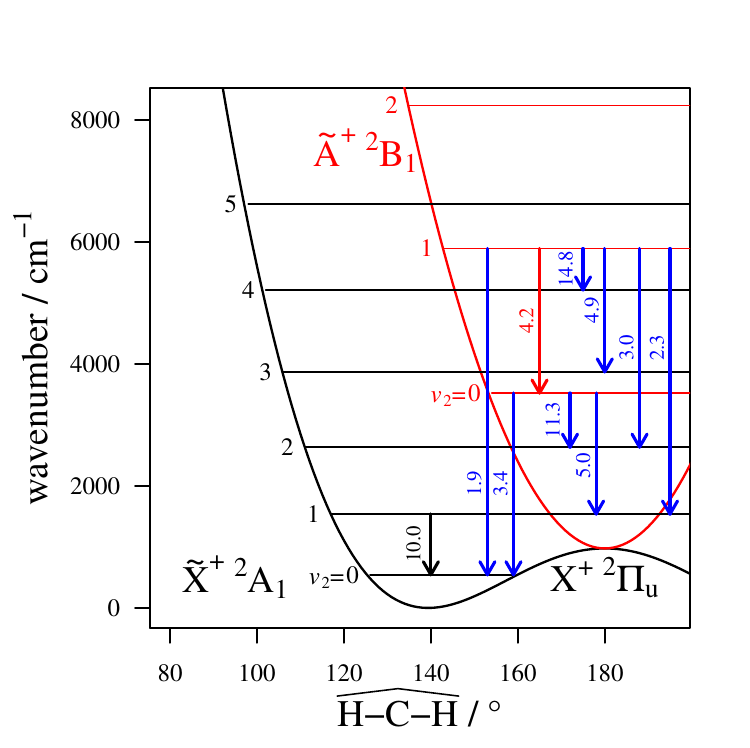}
\caption{Energy-level diagram of the bending levels, with no quanta in either stretching modes, in the two lowest electronic states of CH$_2^+$. These states result from the Renner-Teller coupling, which splits the doubly degenerate $^2\Pi_u$ state into two nondegenerate states for nonlinear configurations. The allowed vibrational (black for the fundamental $\tilde{\rm X}^+\, ^2A_1$ state, red for the $\tilde{\rm A}^+\, ^2B_1$X state) and vibronic (blue) transitions are shown and labeled by their wavelength in $\upmu$m. The bending potentials are plotted as a function of the bending angle $\angle\ce{HCH}$ in degrees. The potential curves and vibrational level positions are adapted from \cite{Jensen1995} and \cite{Coudert2018}.}
\label{Figure_niveaux_ch2_plus}
\end{center}
\end{figure}

\section{${\rm CH}_2^+$ spectroscopy}
\label{sect:spectroscopy}

CH$_2^+$ possesses three vibrational modes: the symmetric stretching mode ($\nu_1$), the bending mode ($\nu_2$), and the asymmetric stretching mode ($\nu_3$). Their corresponding fundamentals in its electronic ground state are located near 2883.0~cm$^{-1}$ \citep[$\sim$3.47 $\upmu$m;][]{Kraemer1994}, 995.5 cm$^{-1}$ \citep[$\sim$10.05 $\upmu$m;][]{Kraemer1994}, and 3131.4 cm$^{-1}$ \citep[$\sim$3.19 $\upmu$m;][]{Rosslein1992}, respectively. 
Although carbocations may behave very differently from the spectroscopic point of view, the observed transitions usually originate from the least energetic vibrational levels within
their electronic ground state when they are detected in space based on their IR emission.
Following the logic behind the detection of CH$^+$ and CH$_3^+$ (observed through their least energetic vibrational levels), the detection of CH$_2^+$ based on the emission from $\nu_2$ would therefore be expected at around 10 $\upmu$m. However, it is very difficult to model the gas-phase emission spectrum of this band accurately. There are no spectroscopic measurements of this band as the available spectroscopic information concerns the rotational constants of the ground state \citep{Rosslein1992, Willitsch2003, Gottfried2004} and the $\nu_3$ band \citep{Wang2013}. In addition to this, the modeling of the $\nu_2$ band is a theoretical challenge \citep{Osmann1997, Bunker2001, Bunker2007, Jensen1995, Coudert2018, Solomonik2008} as the $\nu_2$ mode is a large-amplitude motion and the active mode of a strong vibronic coupling known as the Renner-Teller effect \citep{Herzberg1966,Jungen2019}. Spectroscopic information about the $\nu_2$ band can currently only be retrieved from the theoretical work of \cite{Osmann1997}, who predicted that this band lies between 800 and 1100~cm$^{-1}$.

These results led to the search for CH$_2^+$ emission features between 9 and 11 $\upmu$m.
That CH$_2^+$ was not detected in any of the sources in which CH$^+$ and CH$_3^+$ were detected raises the question why these molecules can be identified by their near- and mid-IR emission fingerprints while the former cannot, even though astrochemical models suggest comparable abundances for CH$_2^+$ and CH$^+$ (see Sect. \ref{sect:PDR_models}). One crucial difference between the CH$^+$ and CH$_3^+$ carbocations and CH$_2^+$ is the close-by electronic state for the latter. In its linear configuration, CH$_2^+$ indeed has a doubly degenerate $\rm^2\Pi_u$ electronic ground state. The aforementioned Renner-Teller effect corresponds to the coupling between the electronic and bending vibrational degrees of freedom ($\nu_2$ mode),
which splits the $\rm ^2\Pi_u$ electronic state into distinct $\tilde{\rm X}^+\, ^2A_1$
and $\tilde{\rm A}^+\, ^2B_1$ nondegenerate states, as illustrated in Figure~\ref{Figure_niveaux_ch2_plus}.
That we do not know how the levels originating from the $\tilde{\rm A}^+\, ^2B_1$  state are populated, hence the importance of the transitions involving these levels in the CH$_2^+$ spectrum, makes it more difficult to validate the consistency with models of the nondetection of CH$_2^+$ based on the signal-to-noise ratio reached by JWST in the 10 $\upmu$m range alone.
The transitions expected in this
range belong to the $\nu_2$ band, and their frequency is altered by the Renner-Teller
coupling. To constrain the detection of the transitions belonging to
this band, we model these transitions in the next sections.

\section{Calculations}
\label{sect:calculations}

The high-resolution spectrum of the CH$_2^+$ molecular ion is difficult to model and constrain for multiple reasons. ($i$) This ion is a highly reactive species that is difficult to produce and study spectroscopically. Consequently, it remains a significant challenge to experimentally constrain the accuracy of the models designed to reproduce its spectroscopic behavior.
({\em ii}) It is a floppy molecule characterized by a low barrier to linearity of about 1000~cm$^{-1}$. Even in its ground vibrational state, \chtwo\ can sample its linear configuration, leading to a theoretically infinite $A$ rotational constant. The resulting strong vibration-rotation coupling leads to quasi-linearity \citep{carter_handy82} and to rotational energies differing from those of an asymmetric top molecule. ({\em iii}) As already stated in Sect.~\ref{sect:spectroscopy}, its $^2\Pi_u$ ground electronic state is split into a lower $\tilde{\rm X}^+\, ^2A_1$ and an upper $\tilde{\rm A}^+\, ^2B_1$ state by the Renner-Teller coupling.

In the approach we used to model the spectrum, the Schr\"odinger equation was
solved simultaneously for all three vibrational modes and
for the overall rotation. The Renner-Teller coupling was taken into account
using the adiabatic representation. The calculation relied on the
tridimensional potential energy surface of \cite{Kraemer1994}
updated by \cite{Jensen1995} to reproduce the $\nu_3$ band
measurements of \cite{Rosslein1992} better. The spin-orbit coupling
was taken into account using the spin-orbit Hamiltonian given
in \cite{Jensen1995}, which was retrieved from the results of \cite{Reuter92}.
The line strengths were computed using the dipole moment surfaces reported
by \cite{Osmann1997}.

Our approach is an extension of two previous calculations. The first calculation, about the photoelectron spectrum of the methylene radical \citep{Coudert2018}, included all the effects described above, except those arising from the asymmetry leading to $|\Delta K| > 0$ matrix elements. The second calculation, dealing with the strong quasi-linearity of the non-rigid FH$_2^+$ ion \citep{gutle12}, accounted exactly for the asymmetry, but there is no Renner-Teller coupling in this species.

The high-resolution spectrum of \chtwo\ has been the subject of several investigations by Bunker and coworkers \citep{Kraemer1994, Osmann1997, Bunker2001}. The investigation of \cite{Osmann1997} was carried out using the same potential, spin-orbit, and dipole moment surfaces as we used here, but a different theoretical treatment based on the MORBID approach \citep{jensen95}. A comparison between the transition wavenumber calculated in this work and the wavenumbers from \cite{Osmann1997} reveals discrepancies of about 0.05~cm$^{-1}$ for pure rotational transitions and of about 0.5~cm$^{-1}$ for transitions belonging to the fundamental bands $\nu_1$, $\nu_2$, and $\nu_3$. For the line strengths, we tend to have small discrepancies, of about 10\%, for strong transitions and larger discrepancies, of about 50\%, for weak transitions. It is unclear whether these discrepancies are due to the different nature of the theoretical approaches or to round-off errors arising when extracting spectroscopic parameters from tables of spectroscopic constants.

The first result of our calculation was a list of energy levels assigned in terms of rotational quantum numbers $N_{K_a K_c}$, vibrational quantum numbers $(v_1v_2v_3)$, electronic states, and electron spin-rotation labels F$_1$ and F$_2$. Denoting $J$ the quantum number for the total angular momentum ${\bf J} = {\bf N}+{\bf S}$, F$_1$ and F$_2$ correspond to ${J}=N+\frac{1}{2}$ and $N-\frac{1}{2}$, respectively. A maximum value of $N$ equal to 10 and an energy cutoff of 13000~cm$^{-1}$ were taken. The $N=0$ level belonging to the ground vibrational state of the $\tilde{\rm X}^+\, ^2A_1$ electronic state was chosen as the zero reference energy. Table~\ref{levs_table} reports low-lying levels with $N \leq 3$ and an energy lower than 1200~cm$^{-1}$. The second result is a line list with assignments, calculated frequencies, line strengths, and Einstein coefficients. For a transition from upper (u) to lower (l) state, the Einstein coefficient $A_{ul}$ in s$^{-1}$ was calculated from the line strength using
\begin{equation}
A_{ul}=\frac{4 (2\pi)^3 10^{-36}\, \nu_{ul}^3 \, S_{ul}}{3\hbar g_u},
\end{equation}
where $\nu_{ul}$ is the wavenumber in cm$^{-1}$, $S_{ul}$ is the strength in Debye$^2$, $\hbar$ is $6.626\,075\,5 \times 10^{-27}/2\pi$, and $g_u$ is the statistical weight of the upper level, excluding statistical nuclear weights. In our case, $g_u=2J_u+1$. Table~\ref{trans_table} shows a portion of the lines list for transitions with $N\leq 5$ and $\nu \leq 60$~cm$^{-1}$. The entire lists of levels and transitions are available via Zenodo.\footnote{\href{https://doi.org/10.5281/zenodo.18892632}{\url{https://doi.org/10.5281/zenodo.18892632}}}

\section{Comparison to JWST observations}
\label{sect:comp_JWST}
Of the sources in which CH$_3^+$ has been detected \citep{Berne2023, Changala2023,Henning2024,Bhatt_2025, Volz2026}, the externally irradiated disk d203-506 observed in the field of view of the PDRs4All observations \citep{Berne2023} has the highest detection in terms of signal-to-noise ratio and is favorable in terms of excitation conditions for CH$_2^+$ to emit in the mid-IR (see Sect. \ref{sect:discussion-detection}). Using the result of the calculations presented in the previous section,
we compared the modeled emission spectra of CH$_2^+$ to the spectrum of d203-506, where CH$^+$ has also been detected.

\begin{figure*}[htbp]
\begin{center}
\includegraphics[width=1.6\columnwidth,angle=0]{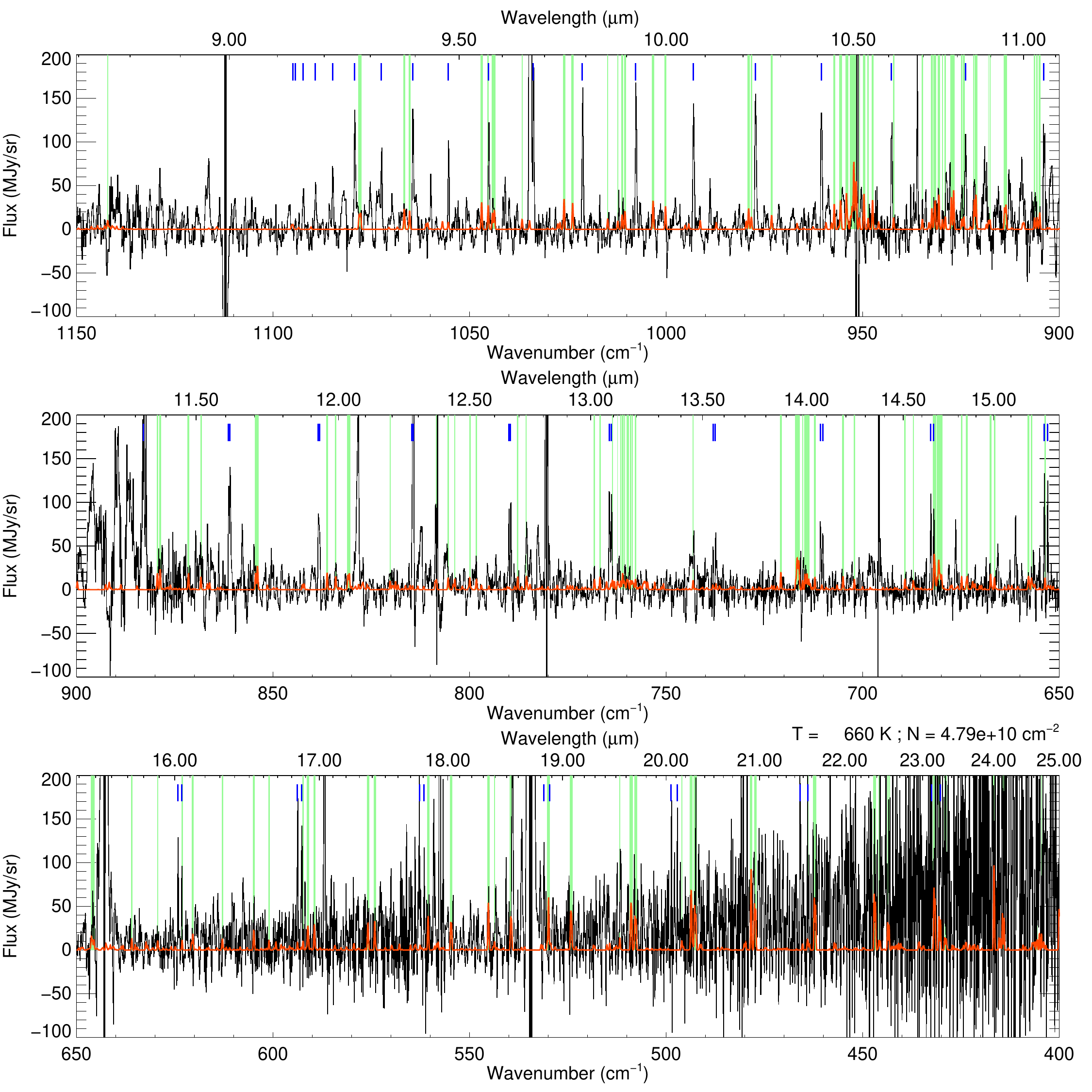}
\caption{Comparison of the irradiated disk d203-506 continuum-subtracted JWST spectrum to the CH$_2^+$ calculated LTE spectrum at 660~K with a column density limit corresponding to $4.8\times10^{10} $~cm$^{-2}$ (red). The series of emission lines labeled with vertical blue marks corresponds to the OH emission reported in \cite{Zannese_2024}. The vertical green shaded regions indicate the CH$_2^+$ lines used 
to constrain the upper limit derived at that excitation temperature.}
\label{Figure_upper_limit_ch2_plus_660K}
\end{center}
\end{figure*}

We used the observations of the MIRI-MRS integral field unit (IFU) mode from the Early Release Science (ERS) program PDRs4All\footnote{\url{https://pdrs4all.org/}, DOI: 10.17909/pg4c-1737}: Radiative feedback from massive stars \citep[ID1288, PIs: Berné, Habart, Peeters,][]{ERS_2022}. The data were reduced using the JWST science pipeline (version 1.17.1) and context jwst\_1322.pmap from the calibration references data system (CRDS; see \cite{Peeters_2024} for the observation parameters and data-reduction process). We used the average spectrum extracted over the apertures given by \cite{Berne2023}. These spectra are in units of $\mathrm{MJy}\,\mathrm{sr}^{-1}$. A smooth spline function with a low spectral frequency continuum was subtracted for the comparison with the modeled spectrum.

Upon close inspection of JWST data of the disk, no obvious emission lines are detected close to the calculated CH$_2^+$ transitions (see Fig. \ref{Figure_upper_limit_ch2_plus_660K}). The typical line position accuracy of the calculations, based on the adopted potential surface, is expected to be better than a wavenumber. Thus, we derived an observational upper limit for the beam-averaged column density of vibrationally excited CH$_2^+$ in d203-506.

The beam-averaged emission from an upper level is given by
\begin{equation}
     J_u = \frac{ h \, \nu_{ul}} {4\pi} A_{ul} \, N_u,
\end{equation}
where $h$ is the Planck constant, $A_{ul}$ is the Einstein coefficient of the considered transition from upper ($u$) to lower ($l$) state, $\nu_{ul}$ is the frequency of the transition, and $N_u$ is the column density of the upper level,
\begin{equation}
N_u = g_{ns}(u) \, [2J(u)+1]  \, {\rm exp}[-h\nu(u)/{kT}] / Q(T),
\end{equation}
with $J(u)$ the total angular momentum, $g_{ns}(u)$ the nuclear spin statistics, $\nu(u)$ the energy of level $u$, and
\begin{equation}
Q(T) = \sum_i g_{ns}(i) \, [2J(i)+1]  \, {\rm exp}[-h\nu(i)/{kT}].
\end{equation}
To estimate the beam-averaged upper limit of an observation for the column density for a given excitation temperature, we interpolated our modeled spectrum after convolution with a Gaussian profile with a wavelength-dependent spectral resolution as reported for the MIRI/MRS instrument \citep{Argyriou2023}, on the observed JWST wavenumber grid. We then selected the channels (i.e., lines) in the modeled spectrum above a threshold of $0.1$ after normalization of the modeled spectrum to its maximum, thus selecting bands in the spectrum that are expected to contribute most to the observed astronomical signal at the selected excitation temperature\footnote{In this method, we considered the lines independently, not in a stacking-like manner.}. We then calculated the weighted $\chi^2(N) = \Sigma_{i}w(i)\times[I_{\rm obs}(i)-I_{\rm model}^N(i)]^2 /\Sigma_{i}w(i)$, where $N$ is the beam-averaged column density of the model, $i$ runs over the spectral channels of the observed spectrum, and $w(i)$ is the weight of this channel, taken as the normalized flux of the modeled spectrum above the threshold value, so that brighter lines have a greater weight. When the modeled column density asymptotically goes to zero, the $\chi^2 (N\rightarrow0)$ asymptotically reaches the noise on the data in the absence of a detection. The upper limit of the column density is then derived when the $\chi^2$ reaches three times this minimum $\chi^2$ (i.e., three times the reduced $\chi^2$) to provide the derived upper limit of the detection.

Fig.~\ref{Figure_upper_limit_ch2_plus_660K} shows a calculated LTE spectrum at 660~K with a column density corresponding to this conservative upper limit on the beam-averaged column density, plotted against the JWST observations in the 1100-400 cm$^{-1}$ ($\sim$9.1--25 $\upmu$m) range, which includes the $\nu_2$ band region. This chosen temperature of 660~K corresponds to the excitation temperature observed for CH$_3^+$ in the disk \citep{Berne2023, Changala2023}.

\begin{figure}
\begin{center}
\includegraphics[width=\columnwidth,angle=0]{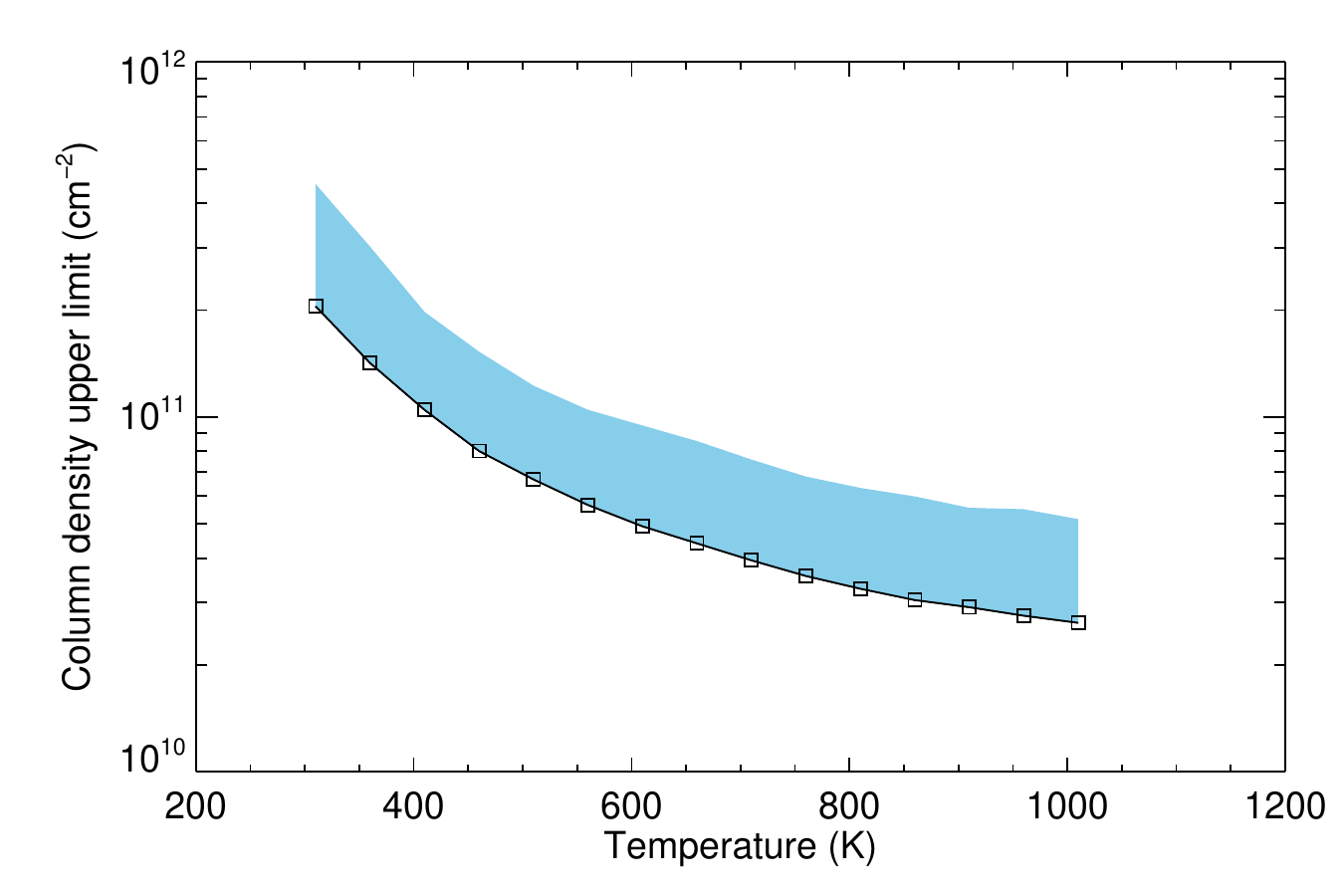}
\caption{Beam-averaged upper limits of the excited column density of CH$_2^+$ for the Orion disk d203-506 derived as a function of the excitation temperature for the calculated model (squares). The blue shaded area is the upper-limit region obtained by applying a Monte Carlo random shift position for each calculated transition (see Sect. \ref{sect:comp_JWST} for details). The excited and non-excited column density in the emitting region is expected to be higher by about 100 times than these values.}
\label{Figure_upper_limit_ch2_plus_T}
\end{center}
\end{figure}

Fig. \ref{Figure_upper_limit_ch2_plus_T} displays the upper limits, depending on the LTE excitation temperature, we derived. 
The upper limit varies with the assumed excitation temperature because the levels are more excited at higher temperatures, and thus, the lines are brighter.  In addition, more lines can be produced, and the probability increases that the lines fall within a region of possible detection.
Except for temperatures well below about 400~K, the upper limits fall below about $\rm 10^{11}$~cm$^{-2}$. To take the potential inaccuracies in the line positions in the model into account, we also made the same calculations using a Monte-Carlo randomization of the transitions around their calculated positions, adding a conservative $\pm$2cm$^{-1}$ deviation, that is, slightly higher than the expected accuracy of the model for the adopted potential energy surface. We calculated 1000 such randomization models for each temperature. The derived upper limits lying above the current adopted model are shown by a shaded area in Fig.~\ref{Figure_upper_limit_ch2_plus_T}. The boundary of the region thus represents the worst-case scenario for line positions with respect to the observed spectrum and is higher by about a factor of two than the current model.

%________________________________________________________

\section{PDR models}
\label{sect:PDR_models}

In this section, we present the results for our fiducial model, which is representative of the external layers of d203-506, from the Meudon PDR code \citep[version 7.1\footnote{Public version of November 2025: \url{https://pdr.obspm.fr/pdr_download.html}},][]{LePetit2006} using similar parameters as for previous models of this source \citep{Berne2023,Goicoechea_2024,Goicoechea_2025_h3+}. We assumed an isochoric model with $n_{\rm H} =10^7$~cm$^{-3}$. We adopted an incident UV field from an illuminating O7 star with an effective temperature $T_{\rm eff}=40000$~K modeled by a blackbody at $T_{\rm eff}$. The distance to the star was set so that the UV field intensity at the ionization front was equal to $G_0=2\times10^4$. We assumed the extinction curve toward HD38087 from \cite{Fitzpatrick_1990} with $R_V = 5.5$. We updated the photodissociation rate of CH$_3^+$ using \cite{Mazo_2024}.

\begin{figure}
\begin{center}
\includegraphics[width=\columnwidth,angle=0]{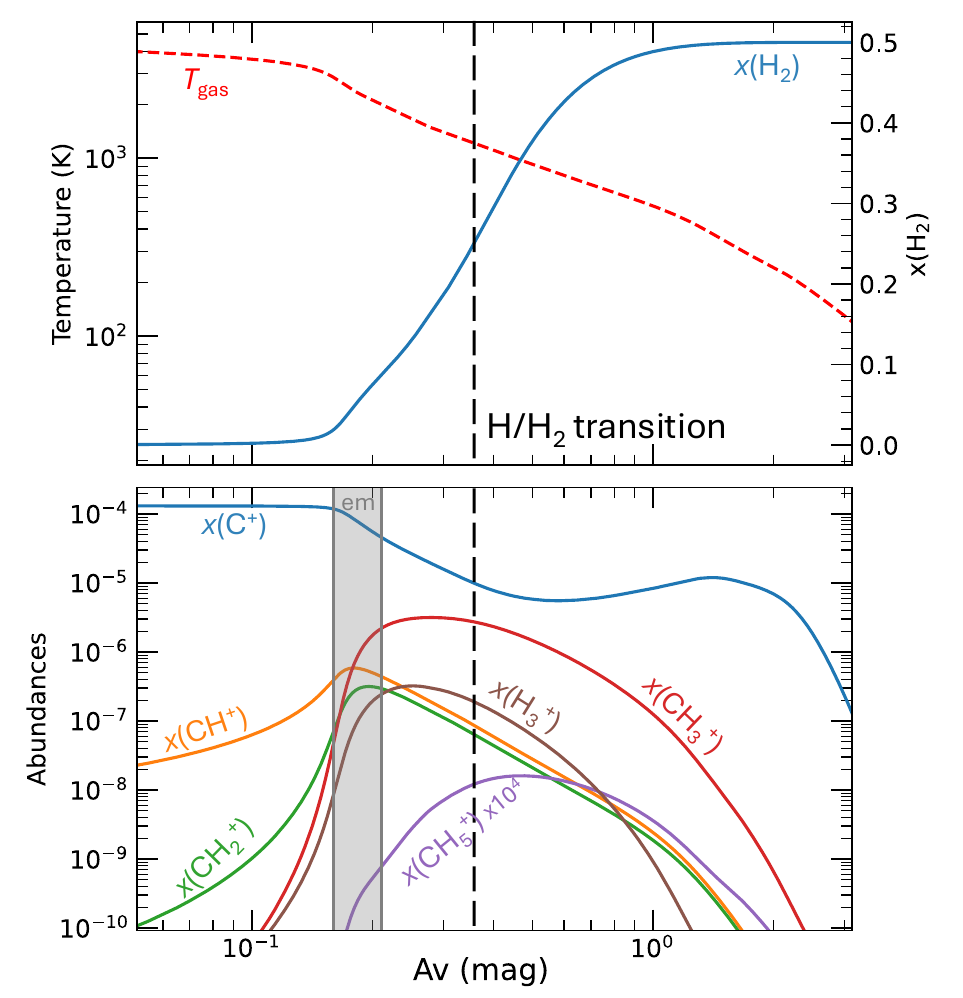}
\caption{Meudon PDR model at $n_{\rm H} = 10^7$~cm$^{-3}$ and $G_0 = 2 \times 10^4$. (Top panel) Temperature and H$_2$ abundance profile as a function of $A_V$. (Bottom panel) Abundances with respect to the proton density of the main species of the carbon chemical chain. The gray area is defined as the emitting region of rovibrational levels of CH$^+$ and CH$_3^+$.}
\label{Figure_modeles}
\end{center}
\end{figure}

\begin{figure}
\begin{center}
\includegraphics[width=\columnwidth,angle=0]{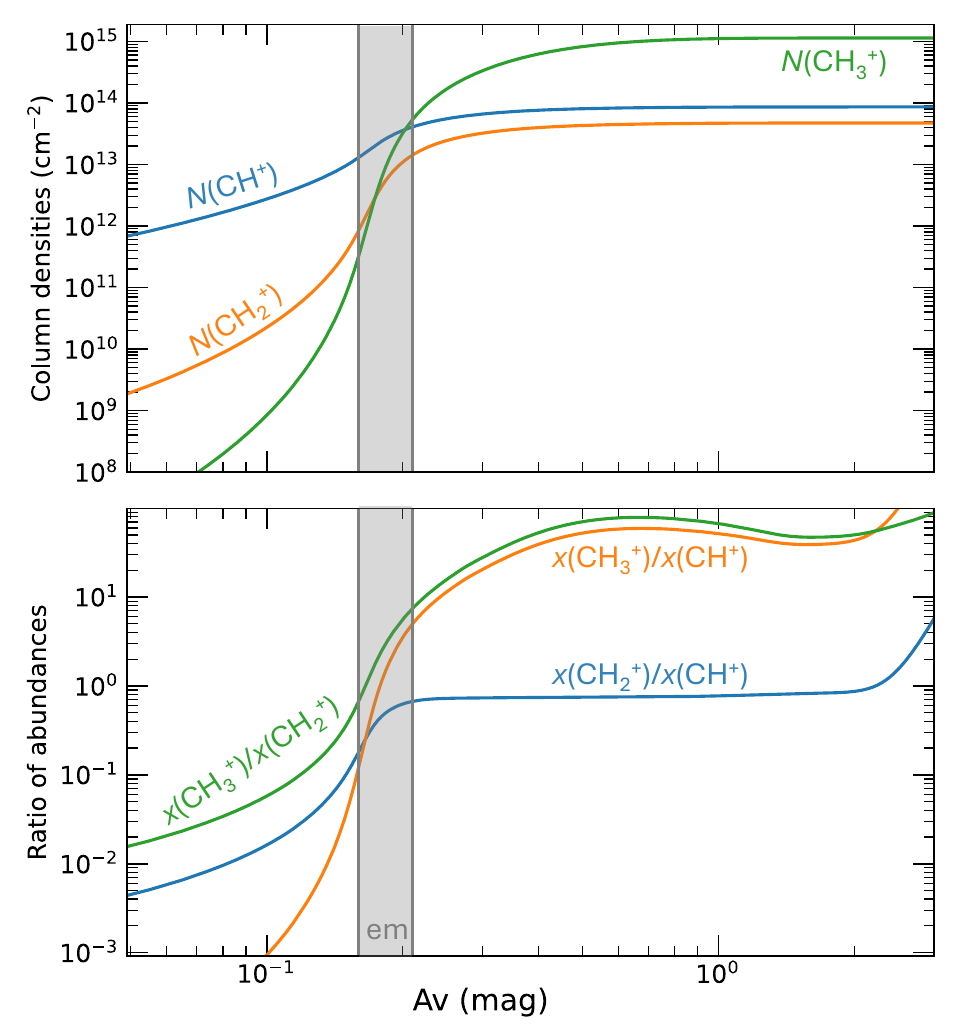}
\caption{Variation in the total column densities and abundances ratio of CH$^+$, CH$_2^+$, and CH$_3^+$ as a function of $A_V$ for a model at $n_{\rm H} = 10^7$~cm$^{-3}$ and $G_0 = 2 \times 10^4$. The gray area is defined as the emitting region for rovibrational transitions of CH$^+$ and CH$_3^+$. }
\label{Figure_modeles_ratio}
\end{center}
\end{figure}

Figure \ref{Figure_modeles} shows the temperature and abundance profile of the studied species as a function of depth inside the PDR (in units of extinction in the visible). This figure shows that CH$^+$, CH$_2^+$, and CH$_3^+$ peak at roughly the same position (near the H/H$_2$ transition, where $x(\rm H_2) = 0.25$). This model predicts the CH$_3^+$ abundance to be higher by an order of magnitude than CH$^+$ and CH$_2^+$ at the dissociation front. CH$^+$ and CH$_2^+$ have a very similar abundance at this position, with CH$^+$ being slightly more abundant (by a factor of 2). Even though CH$^+$ peaks at the same position as CH$_2^+$ and CH$_3^+$, its abundance is higher closer to the edge of the PDR due to the need for excited H$_2$ to overcome the endothermicity of the C$^+$ + H$_2$ = CH$^+$ + H reaction. In contrast, the CH$_3^+$ abundance decreases more slowly deeper into the PDR because it is not efficiently destroyed by reactions with H$_2$ such as CH$^+$ and CH$_2^+$. Fig. \ref{Figure_modeles_ratio} highlights these variations in morphology in the model by displaying the column density profiles and the variation in the abundance ratios as a function of $A_V$. 
Our fiducial model predicts $N($CH$^+) = 9.2 \times 10^{13}$~cm$^{-2}$, $N($CH$_2^+) = 5.3 \times 10^{13}$~cm$^{-2}$, and  $N($CH$_3^+) = 1.1 \times 10^{15}$~cm$^{-2}$. Hence, the model predicts the column density of CH$_2^+$ to be twice lower than that of CH$^+$ and 20 times lower than that of CH$_3^+$. The fraction of excited species observed, in the case of CH$^+$ and CH$_3^+$, is lower than these predictions, but with similar column densities of $N(\rm CH^+) = (3.3 \pm 0.3)\times10^{11}$~cm$^{-2}$ and $N(\rm CH_{3}^{+}) = (5.4 \pm3.2)\times10^{11}$~cm$^{-2}$ observed in the rovibrational states. This means that the column density ratio of the rovibrationally excited (observed) to total population (modeled) is different for CH$^+$ and CH$_3^+$, around 0.3$\%$ for CH$^+$ and 0.05$\%$ for CH$_3^+$. However, Fig. \ref{Figure_modeles} reveals that CH$_3^+$ is more abundant at high $A_V$, at deeper positions than where the emission of rovibrational levels is expected to originate. The emitting region of rovibrational levels is expected around the abundance peak of CH$^+$ and CH$_2^+$. We can define the emitting region (later named "em", represented as the shaded area in Fig. \ref{Figure_modeles} and \ref{Figure_modeles_ratio}) where the abundance of CH$^+$ is at least 75\% of its maximum abundance, which corresponds to $A_V = 0.16-0.21$. Then, in this region $N_{\rm em}($CH$^+) \sim 3 \times 10^{13}$~cm$^{-2}$ and $N_{\rm em}($CH$_3^+) \sim 6 \times 10^{13}$~cm$^{-2}$. The fraction of excited levels is similar and about 1$\%$. Hence, we also expect that for d203-506, the column density of rovibrational levels of CH$_2^+$ is about 1$\%$ of its column density at low $A_V$. In the fiducial model, $N_{\rm em}($CH$_2^+) \sim 1.5 \times 10^{13}$~cm$^{-2}$, so we expect $N_{\rm exc}($CH$_2^+) \sim 1.5 \times 10^{11}$~cm$^{-2}$. In the emitting region, the column density of these species varies drastically (see Fig. \ref{Figure_modeles_ratio}). The percentage of the excited column density might therefore be lower or higher than 1$\%$ and should be treated as a rough estimate.

Depending on the underlying excitation temperature, this is above or just below the upper limit we inferred from the comparison to JWST spectra shown in Fig.~\ref{Figure_upper_limit_ch2_plus_T}. If CH$_2^+$ has an equivalent excitation temperature of about 600~K as observed for CH$_3^+$, we should still see hints of detection of some well-isolated transitions at this level, such as the doublet around 1025 cm$^{-1}$ (see Fig.~\ref{Figure_upper_limit_ch2_plus_660K}). These models suggest that pushing the JWST integration time might reveal the presence of CH$_2^+$, as it currently seems to be at the detection limit.

\section{Discussion}

\subsection{Detection of CH$_2^+$}

\label{sect:discussion-detection}

Since it is a very reactive intermediate in the ion reaction network, CH$_2^+$ is difficult to observe. Nevertheless, constraining its abundance provides insight into chemical networks, as models expect it to be only several times less abundant than CH$_3^+$ and CH$^+$ in peak excitation regions.
Assuming its excitation temperature to be of the same order as the one observed for most species detected in the near- and mid-IR in the Orion Bar with JWST, spanning the 600-1000~K range for most of them, we estimated a CH$_2^+$ excited column density upper limit that is lower by about one order of magnitude than the one detected for CH$_3^+$.  
We also provide a list of transitions for calculating CH$_2^+$ spectra and for hopefully obtaining a detection or more constraining upper limits from future observations, thus providing constraints for astrochemical models.
Regions that therefore should favor CH$_2^+$ are those in which it is less efficiently destroyed, that is, those with a relatively low H$_2$ density (but still high enough for it to form), short dynamical timescales, and a high ionization fraction. Conversely, regions that favor CH$_3^+$ include a higher abundance of H$_2$ and higher densities, which would increase the efficiency of the carbocation reaction chain, described in reactions (\ref{eq:C+}), (\ref{eq:H3+}). This difference in required conditions is observed in PDR models, where the CH$_3^+$/CH$_2^+$ ratio decouples at moderate $A_V$ (see Fig. \ref{Figure_modeles_ratio}). In addition, CH$_3^+$ has a lower electronic recombination efficiency, and thus, quasi-stationary conditions will favor it.
When CH$^+$ is formed, conversion into CH$_2^+$ by interaction with H$_2$ is rapid and exothermic. Then, it proceeds rapidly to form CH$_3^+$ and CH$_3^+$ appears as the relaxed product of any injection of CH$^+$, CH$_2^+$ being an intermediate.
The ratios CH$^+$/CH$_2^+$/CH$_3^+$ should therefore measure the equilibrium between energy injection (endoergic formation for CH$^+$) and chemical relaxation (hydrogenation towards CH$_3^+$), which makes them direct tracers of out-of-equilibrium energetic dissipation of energy in interstellar gas.

\subsection{The H$_2$ + CH$^+$ $\longrightarrow$ CH$_2^+(\widetilde X,\widetilde A)$ +H formation reaction}
\label{sect:reaction}
A deepest understanding of CH$_2^+$ formation pathways is required, in addition to its spectroscopic properties, to assess whether the nondetection of CH$_2^+$ is compatible with current predictions by astrochemical models.  First, it is important to determine the population distribution of the $\rm CH_2^+$ energy levels following its formation via H$_2$ + CH$^+$, which might deviate from an LTE-like distribution at temperatures lower than 1000~K. $\rm CH_2^+$ might also be formed in very excited $\nu_2$ levels in the $\rm \tilde{X}$ state or even in its first electronic excited state ($\rm \tilde{A}^+$).

\begin{figure}
    \center
    \includegraphics[scale=0.45,trim={2cm 6cm 2cm 6cm},clip]{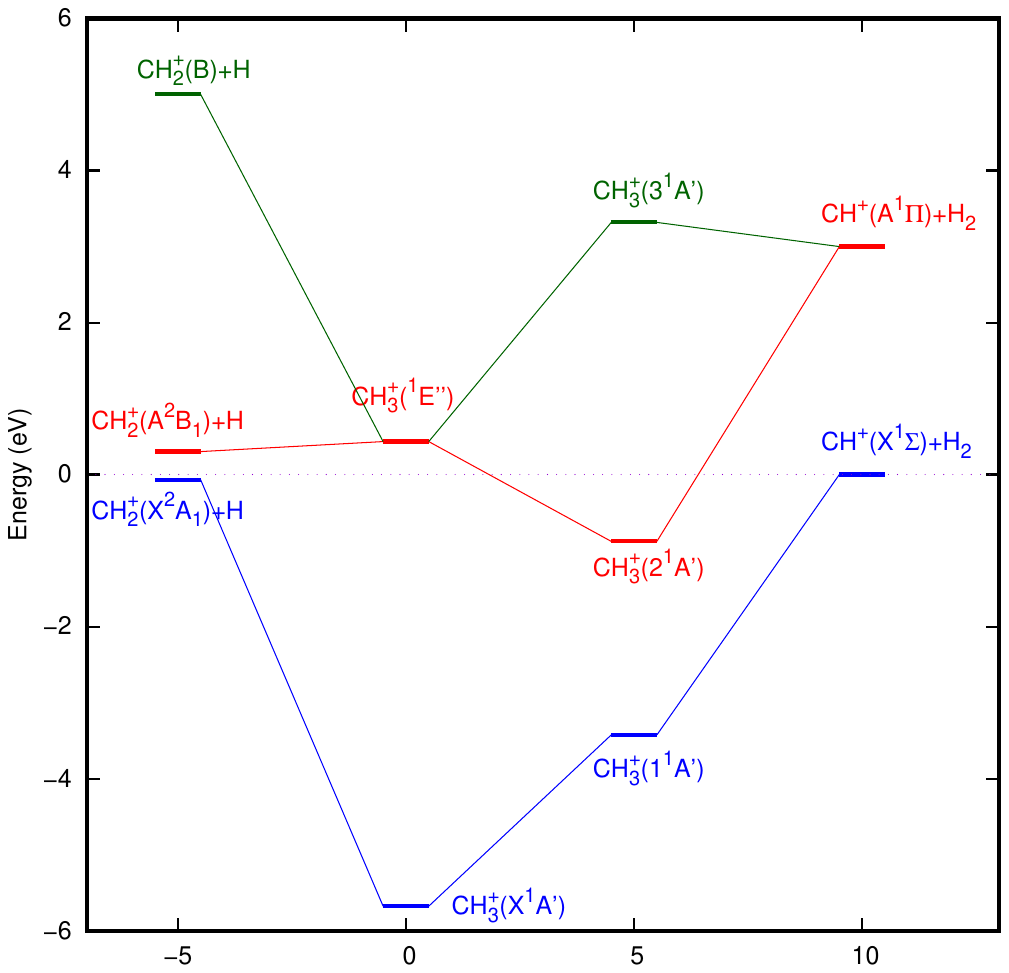}
    \caption{Energy diagram of the lower electronic states involved in the
      H$_2$ + CH$^+$ $\longrightarrow$ CH$_2^+$ + H  reaction from \cite{Mazo_2024}
   , including zeropoint energies.}
    \label{fig:meps}
\end{figure}

The reaction
\begin{eqnarray}
  {\rm CH}^+(X^1\Sigma) + {\rm H}_2(X^1\Sigma^+) \rightarrow {\rm CH}^+_2 (\widetilde X ^2A_1, \widetilde A ^2B_1) + {\rm H}(^2S)
\end{eqnarray}
proceeds through the formation of the CH$_3^+(\widetilde X^1A')$ system and is slightly exothermic.
The ground and first two excited electronic states of the full CH$_3^+$ system were calculated recently
for the study of CH$_3^+(\widetilde X^1A')$ photodissociation \citep{Mazo_2024}, and the
energy diagram with some stationary points (including zeropoint energy) is shown in Fig.~\ref{fig:meps}.
Clearly, the CH$_2^+ (\widetilde X ^2A_1)$ correlates adiabatically with the reactants, in a reaction
that is exothermic by only $\approx$ 66 meV, using the explicitly correlated
  internally contracted multireference configuration interaction (ic-MRCI-F12)  method
  \citep{Werner-Knowles:88,Shiozaki-Werner:11}, with the MOLPRO suite
  of programs \citep{MOLPRO-WIREs} and the cc-pCVTZ-F12 electronic basis set \citep{Hill-etal:10}.
  
  In contrast, the formation of the excited
CH$_2^+ (\widetilde A ^2B_1)$ products correlates with the electronically excited CH$^+(A^1\Pi)$.
Transitions among these two electronic states are negligible along the reaction path until they form
the CH$_2^+$ products, where the energy separation is significantly reduced. The symmetry of the  two electronic states
of CH$_2^+$ is different, and they are only coupled by the weak spin-orbit and electronic Coriolis force, which cause the Renner-Teller effects.
These couplings are important for accurately describing the spectroscopy of the ions, but they are rather
small, so the electronic transitions between them can be neglected. Moreover, the reaction to form excited
CH$_2^+ (\widetilde A ^2B_1)$ is endothermic by $\approx$ 303  meV,
so that at low temperatures, this reactive channel is unreachable.

In summary, in the H$_2$ + CH$^+$ reaction mainly CH$_2^+(\widetilde X)$ is expected to be formed, following a statistical mechanism discussed in Appendix \ref{appendix:CH2+_rate}, with a rate of 1.01 10$^{-9}$ (T/300)$^{-0.193}$ exp(-16.0/T) cm$^3$s$^{-1}$ for 20 $<T<$ 1000 K. In the KIDA database, a constant value of 
1.20 10$^{-9}$ cm$^3$s$^{-1}$  is reported, based on a estimation by \cite{Anicich_1993}. Because of the statistical mechanism, the 
CH$_2^+(\widetilde X)$ is expected to be formed with a rovibrational population following a Boltzmann distribution.

\subsection{Importance of transitions from the $\tilde{\rm A}^+$ level}

If CH$_2^+(\widetilde A)$ is only a minor product in the reaction CH$^+$ + H$_2$, it can be formed in this excited electronic state through alternative routes. One of these routes is the photodissociation of CH$_3^+({\widetilde X})$, as recently reported by \cite{Mazo_2024} and displayed in the energy diagram of Fig.~\ref{fig:meps}. If this route becomes important, $\rm CH_2^+$ might deexcite through a vibrational transition from its least energetic vibrational mode ($\rm\nu_2=1 \rightarrow \nu_2=0$ in $\rm \tilde{A}^+$), which is expected around 2500 cm$^{-1}$ ($\sim 4$ $\upmu$m). Alternatively, it might relax to its ground electronic state ($\rm \tilde{X}^+$) via vibronic emission. As demonstrated by \cite{Wang2013}, the large change in geometry between the $\rm \tilde{X}^+$ and $\rm \tilde{A}^+$ states implies a strong nondiagonal Franck-Condon progression along the $\rm\nu_2$ vibrational mode in the $\rm \tilde{A}^+ \leftarrow \tilde{X}^+$ absorption spectrum of $\rm CH_2^+$. The emission spectrum from $\rm \tilde{A}^+$ state toward $\rm \tilde{X}^+$ will thus also present a long vibrational progression, but at a lower frequency (or higher wavelength). If only the first vibrational levels of the $\rm \tilde{A}^+$ state are populated, based on the CH$^+$ and CH$_3^+$ excitation temperatures in the different objects in which it was detected (between 700 and 1500 K), the strongest emission transitions will be toward the excited vibrational levels of $\rm \tilde{X}^+$. We would then expect emission features of $\rm CH_2^+$ from the $\rm \nu_2=0$ and $\rm \nu_2=1$ of the $\rm \tilde{A}^+$ electronic state toward the $\rm \nu_2$ bending levels of the $\rm \tilde{X}^+$ electronic state. The expected ranges for such transitions are reported in Fig~\ref{Figure_niveaux_ch2_plus}. Typical vibronic emissions present a higher probability than vibrational emissions, and CH$_2^+$ spectroscopic signatures might therefore be spread (large number of transitions) and contribute to other spectral regions. The emission spectrum of $\rm CH_2^+$ in hot environments might even be located in the near-IR region, especially if $\rm CH_2^+$ is produced in vibrational states higher than $\rm \nu_2$=1 in the $\rm \tilde{A}^+$ electronic state. Further investigation of $\rm CH_2^+$ formation pathways and degree of excitation is thus required.
\section{Conclusion}
We investigated the observational conditions required to constrain the presence of the particularly hard-to-detect CH$_2^+$ reactive species. We calculated and identified mid- and far-IR transitions expected to be most relevant, with an emphasis on low-lying rovibrational states. Synthetic emission spectra were produced for a range of excitation temperatures and compared with observations of the irradiated disk d203-506, in which chemically related species are known to be present. The main conclusions of this study are summarized below
\begin{enumerate}
    \item We derived observationally temperature-dependent upper limits on the CH$_2^+$ excited column density for the irradiated disk d203-506, which were found to be significantly lower (about an order of magnitude) than the column densities measured for other carbocations in excited states.
    \item Using a fiducial PDR model of d203-506, we showed that CH$^+$, CH$_2^+$, and CH$_3^+$ are expected to originate from the same region, near the H/H$_2$ transition, and to have very similar abundances.
    \item This model predicts that the column density of rovibrational levels of CH$^+$ and CH$_3^+$ only represents $1\%$ of their column density in the emitting region. Assuming that CH$_2^+$ follows the same logic, we predict a column density that is consistent with its nondetection in d203-506.
    \item The CH$_3^+$ and CH$_2^+$ ratio is expected to decouple at moderate $A_V$, which indicates that the regions that are expected to favor CH$_2^+$ are those with a relatively low  H$_2$ density, short dynamical timescales, and a high ionization fraction. The ratios CH$^+$/CH$_2^+$/CH$_3^+$ are expected to measure the equilibrium between energy injection and chemical relaxation.
\end{enumerate}

Future laboratory measurements are essential for benchmarking and refining the theoretical predictions.
While we await these laboratory experiments to evaluate the accuracy of the calculations, and thus, to refine the astronomical search for CH$_2^+$, we present a comprehensive set of tabulated transitions to support future observational searches for this carbocation. These data will facilitate either direct detections or the derivation of more stringent upper limits, thereby providing critical benchmarks for astrochemical models, particularly in photon-dominated environments.

\begin{acknowledgements}
\newline

This work is based [in part] on observations made with the NASA/ESA/CSA James Webb Space Telescope. The data were obtained from the Mikulski Archive for Space Telescopes at the Space Telescope Science Institute, which is operated by the Association of Universities for Research in Astronomy, Inc., under NASA contract NAS 5-03127 for JWST. These observations are associated with program \#1288 (DOI: 10.17909/pg4c-1737).
Support for program \#1288 was provided by NASA through a grant from the Space Telescope Science Institute, which is operated by the Association of Universities for Research in Astronomy, Inc., under NASA contract NAS 5-03127, and the Canadian Space Agency (CSA, 22JWGO1-16). This work was performed in part at the French MIRI center with the support of CNES and the ANR-labcom INCLASS between IAS and ACRI-ST, and also supported by the Program National “Physique et Chimie du Milieu Interstellaire” (PCMI) of CNRS/INSU with INC/INP cofunded by CEA and CNES. M.Z. and J.R.G. thank the Spanish MCINN for funding support under grant PID2023-146667NB-I00. M. Z. acknowledges the Juan de la Cierva Postdoctoral Fellow project JDC2024-054658-I, funded by MICIU/AEI/10.13039/501100011033 and by the ESF+. O.R. and P.dM-D thank Spanish MCINN for funding support under grant PID2024-156686NB-I00. E.P. and J.C. acknowledge support from the University of Western Ontario, the Institute for Earth and Space Exploration, the Canadian Space Agency (CSA, 22JWGO1-16), and the Natural Sciences and Engineering Research Council of Canada. C.B. acknowledges support from the Internal Scientist Funding Model (ISFM) Laboratory Astrophysics Directed Work Package Round 3 at NASA Ames and is grateful for an appointment at NASA Ames Research Center through the San José State University Research Foundation (80NSSC22M0107).

\end{acknowledgements}

\bibliographystyle{aa}

\appendix 
\FloatBarrier
\nolinenumbers
\section{CH$_2^+({\widetilde X})$ formation rate}

\label{appendix:CH2+_rate}
The ground state potential is attractive  between CH$^+$  and H$_2$, with no barrier to form CH$_3^+$,
so that the reaction takes place through the formation of long-lived (CH$_3^+$)$^*$ complexes, followed by the dissociation of the complex either towards CH$^+$+H$_2$ or to CH$_2^+({\widetilde X})$ + H, channels. 

The CH$^+$+H$_2$ $\longrightarrow$CH$_2^+({\widetilde X})$ + H micro-canonical reaction rate constant is given by
\begin{eqnarray}\label{eq:reaction-rate}
K(E)= \sigma_c(E) \frac{K_{CH_2^++H}(E)}{K_{CH_2^++H}(E)+ K_{CH^++H_2}(E)}
\end{eqnarray}
The capture cross section, $\sigma_c(T)=\pi \sqrt{e^2 2\alpha /E_t}$ for the formation
of these complexes is dominated by the charge-induced electric dipole long-range interaction, giving a Langevin
rate for hydrogen molecule of  2.25 10$^{-9}$ cm$^3$s$^{-1}$ using the H$_2$ polarizability of \cite{Woon-Herbst:09}.

The unimolecular dissociation rate constants are calculated using the statistical RRKM theory \citep{Marcus:52,Marcus:52b}. The microcanonical rate constants to dissociate the (CH$_3^+$)$^*$ complex are evaluated towards each re-arrangement channel $\alpha$ (CH$^+$+H$_2$ or CH$_2^+$ + H) independently at each total energy $E$ and total angular momentum $J$ as \citep{Marcus:52,Marcus:52b,Miller:79,Miller:87}
\begin{eqnarray}\label{eq:rrkm-rate}
K_\alpha(E,J)= \frac{\sum_{q,\ell} P(E, J, q,\ell)}{2 \pi \hbar \rho(E,J)}
\end{eqnarray}
where $q= v_A,v_B,J_A,J_B,K_A,K_B$ denotes the quantum numbers defining the states of fragments A and B, with eigenvalues $\epsilon_{q}$ 
and $\rho(E,J)$ is the density of states of the  (CH$_3^+$)$^*$ complex. In the above expression, the probability is evaluated classically as $P(E, J, q,\ell) = 1$ for $E>E^b_{q\ell}$ (and zero elsewhere), with $E^b_{q\ell}$ being the rotational barrier.
In all these expressions, the rovibrational eigenvalues are evaluated in a rigid rotor approximation of harmonic oscillators for H$^+$, H$_2$ , CH$_2^+$ and CH$_3^+$ at their corresponding equilibrium geometries. 
The corresponding reaction rate of Eq.~\ref{eq:reaction-rate} is then integrated numerically  to obtain the thermal reaction rate constant reported in section \ref{sect:reaction}.

\section{Calculated tables of levels}
Table~\ref{levs_table} shows a small portion of the level list
computed in Section~\ref{sect:calculations}.

\begin{table}[!h]
\begin{center}
\caption{\label{levs_table}A portion of the levels computed
in Section~\ref{sect:calculations}.}
\begin{tabular}{ccccc c c}\hline
	$N_{K_a K_c}$ & F$_{12}$ & $v$ & $L$ & $\Gamma$ & $E(\mbox{cm$^{-1}$})$ & $g$ \\ \hline
	$0_{00}$ & F$_1$ & (000) & $\tilde{X}^+$ & $A_1$ & \spcn{3}0.000 & 1 \\
	$1_{01}$ & F$_2$ & (000) & $\tilde{X}^+$ & $B_1$ & \spcn{2}14.692 & 3 \\
	$1_{01}$ & F$_1$ & (000) & $\tilde{X}^+$ & $B_1$ & \spcn{2}14.692 & 3 \\
	$2_{02}$ & F$_2$ & (000) & $\tilde{X}^+$ & $A_1$ & \spcn{2}44.062 & 1 \\
	$2_{02}$ & F$_1$ & (000) & $\tilde{X}^+$ & $A_1$ & \spcn{2}44.062 & 1 \\
	$1_{10}$ & F$_2$ & (000) & $\tilde{X}^+$ & $A_2$ & \spcn{2}74.694 & 1 \\
	$1_{11}$ & F$_2$ & (000) & $\tilde{X}^+$ & $B_2$ & \spcn{2}75.528 & 3 \\
	$1_{11}$ & F$_1$ & (000) & $\tilde{X}^+$ & $A_2$ & \spcn{2}77.113 & 1 \\
	$1_{10}$ & F$_1$ & (000) & $\tilde{X}^+$ & $B_2$ & \spcn{2}77.951 & 3 \\
	$3_{03}$ & F$_2$ & (000) & $\tilde{X}^+$ & $B_1$ & \spcn{2}88.082 & 3 \\
	$3_{03}$ & F$_1$ & (000) & $\tilde{X}^+$ & $B_1$ & \spcn{2}88.082 & 3 \\
	$2_{11}$ & F$_2$ & (000) & $\tilde{X}^+$ & $B_2$ & \spcn{1}104.064 & 3 \\
	$2_{11}$ & F$_1$ & (000) & $\tilde{X}^+$ & $B_2$ & \spcn{1}105.395 & 3 \\
	$2_{12}$ & F$_2$ & (000) & $\tilde{X}^+$ & $A_2$ & \spcn{1}106.567 & 1 \\
	$2_{12}$ & F$_1$ & (000) & $\tilde{X}^+$ & $A_2$ & \spcn{1}107.904 & 1 \\
	$3_{12}$ & F$_2$ & (000) & $\tilde{X}^+$ & $A_2$ & \spcn{1}147.127 & 1 \\
	$3_{13}$ & F$_1$ & (000) & $\tilde{X}^+$ & $A_2$ & \spcn{1}148.058 & 1 \\
	$3_{13}$ & F$_2$ & (000) & $\tilde{X}^+$ & $B_2$ & \spcn{1}152.126 & 3 \\
	$3_{12}$ & F$_1$ & (000) & $\tilde{X}^+$ & $B_2$ & \spcn{1}153.065 & 3 \\
	$2_{20}$ & F$_2$ & (000) & $\tilde{X}^+$ & $B_1$ & \spcn{1}310.574 & 3 \\
	$2_{21}$ & F$_2$ & (000) & $\tilde{X}^+$ & $A_1$ & \spcn{1}310.581 & 1 \\
	$2_{20}$ & F$_1$ & (000) & $\tilde{X}^+$ & $B_1$ & \spcn{1}313.774 & 3 \\
	$2_{21}$ & F$_1$ & (000) & $\tilde{X}^+$ & $A_1$ & \spcn{1}313.781 & 1 \\
	$3_{22}$ & F$_2$ & (000) & $\tilde{X}^+$ & $A_1$ & \spcn{1}355.508 & 1 \\
	$3_{21}$ & F$_2$ & (000) & $\tilde{X}^+$ & $B_1$ & \spcn{1}355.543 & 3 \\
	$3_{22}$ & F$_1$ & (000) & $\tilde{X}^+$ & $A_1$ & \spcn{1}357.746 & 1 \\
	$3_{21}$ & F$_1$ & (000) & $\tilde{X}^+$ & $B_1$ & \spcn{1}357.781 & 3 \\
	$3_{31}$ & F$_2$ & (000) & $\tilde{X}^+$ & $B_2$ & \spcn{1}649.744 & 3 \\
	$3_{30}$ & F$_2$ & (000) & $\tilde{X}^+$ & $A_2$ & \spcn{1}649.744 & 1 \\
	$3_{31}$ & F$_1$ & (000) & $\tilde{X}^+$ & $A_2$ & \spcn{1}653.263 & 1 \\
	$3_{30}$ & F$_1$ & (000) & $\tilde{X}^+$ & $B_2$ & \spcn{1}653.264 & 3 \\
	$1_{10}$ & F$_2$ & (010) & $\tilde{X}^+$ & $A_2$ & \spcn{1}956.121 & 1 \\
	$1_{11}$ & F$_2$ & (010) & $\tilde{X}^+$ & $B_2$ & \spcn{1}956.829 & 3 \\
	$1_{10}$ & F$_1$ & (010) & $\tilde{X}^+$ & $A_2$ & \spcn{1}966.662 & 1 \\
	$1_{11}$ & F$_1$ & (010) & $\tilde{X}^+$ & $B_2$ & \spcn{1}967.382 & 3 \\
	$2_{12}$ & F$_2$ & (010) & $\tilde{X}^+$ & $B_2$ & \spcn{1}987.859 & 3 \\
	$2_{11}$ & F$_2$ & (010) & $\tilde{X}^+$ & $A_2$ & \spcn{1}989.990 & 1 \\
	$2_{11}$ & F$_1$ & (010) & $\tilde{X}^+$ & $B_2$ & \spcn{1}993.698 & 3 \\
	$0_{00}$ & F$_1$ & (010) & $\tilde{X}^+$ & $A_1$ & \spcn{1}995.655 & 1 \\
	$2_{12}$ & F$_1$ & (010) & $\tilde{X}^+$ & $A_2$ & \spcn{1}995.848 & 1 \\
	$1_{01}$ & F$_1$ & (010) & $\tilde{X}^+$ & $B_1$ & 1010.339 & 3 \\
	$1_{01}$ & F$_2$ & (010) & $\tilde{X}^+$ & $B_1$ & 1010.339 & 3 \\
	$3_{13}$ & F$_2$ & (010) & $\tilde{X}^+$ & $A_2$ & 1031.304 & 1 \\
	$3_{12}$ & F$_1$ & (010) & $\tilde{X}^+$ & $A_2$ & 1035.392 & 1 \\
	$3_{12}$ & F$_2$ & (010) & $\tilde{X}^+$ & $B_2$ & 1035.562 & 3 \\
	$3_{13}$ & F$_1$ & (010) & $\tilde{X}^+$ & $B_2$ & 1039.676 & 3 \\
	$2_{02}$ & F$_1$ & (010) & $\tilde{X}^+$ & $A_1$ & 1039.695 & 1 \\
	$2_{02}$ & F$_2$ & (010) & $\tilde{X}^+$ & $A_1$ & 1039.695 & 1 \\
	$3_{03}$ & F$_2$ & (010) & $\tilde{X}^+$ & $B_1$ & 1083.697 & 3 \\
	$3_{03}$ & F$_1$ & (010) & $\tilde{X}^+$ & $B_1$ & 1083.698 & 3 \\
	\hline\hline
\end{tabular}

\tablefoot{Levels are assigned using
rotational quantum numbers $N_{K_a K_c}$, the electron spin-rotation
label F$_{12}$, the vibrational quantum numbers $v=(v_1v_2v_3)$,
the electronic state label $L$,
and their $C_{2v}$ symmetry $\Gamma$. Energies
in cm$^{-1}$ and nuclear spin statistical weights are given in the column
headed $E$ and $g$, respectively.}
\end{center}
\vspace{-5cm}
\end{table}

\onecolumn
\section{Calculations details}
Table~\ref{trans_table} shows a small portion of line list calculated in Section~\ref{sect:calculations}.

\begin{table*}[!h]
\begin{center}
\caption{\label{trans_table}A portion of the transitions computed
in Section~\ref{sect:calculations}.}
\begin{tabular}{cccc c cccc c c c}\hline
	\multicolumn{4}{@{}c@{}}{Upper level} & & \multicolumn{4}{@{}c@{}}{Lower level} \\ \cline{1-4} \cline{6-9}
	$N_{K_a K_c}$ & F$_{12}$ & $v$ & $L$ & & $N_{K_a K_c}$ & F$_{12}$ & $v$ & $L$ & $\nu$ & $S$ & $A_{12}$ \\ \hline
	$3_{13}$ & F$_1$ & (000) & $\tilde{X}^+$ & & $4_{04}$ & F$_1$ & (000) & $\tilde{X}^+$ & \spcn{1}1.34912 & 0.702787e$+$0 & 0.676538e$-$7 \\
	$4_{14}$ & F$_1$ & (010) & $\tilde{X}^+$ & & $3_{03}$ & F$_1$ & (010) & $\tilde{X}^+$ & \spcn{1}7.69236 & 0.800667e$+$0 & 0.114297e$-$4 \\
	$2_{02}$ & F$_2$ & (010) & $\tilde{X}^+$ & & $3_{13}$ & F$_2$ & (010) & $\tilde{X}^+$ & \spcn{1}8.39151 & 0.457150e$+$0 & 0.211798e$-$4 \\
	$5_{05}$ & F$_2$ & (000) & $\tilde{X}^+$ & & $4_{13}$ & F$_1$ & (000) & $\tilde{X}^+$ & 14.88201 & 0.173483e$-$1 & 0.179326e$-$5 \\
	$5_{05}$ & F$_1$ & (000) & $\tilde{X}^+$ & & $4_{13}$ & F$_1$ & (000) & $\tilde{X}^+$ & 14.88329 & 0.936534e$+$0 & 0.806940e$-$4 \\
	$5_{05}$ & F$_2$ & (000) & $\tilde{X}^+$ & & $4_{14}$ & F$_2$ & (000) & $\tilde{X}^+$ & 15.60108 & 0.762806e$+$0 & 0.908406e$-$4 \\
	$2_{11}$ & F$_2$ & (000) & $\tilde{X}^+$ & & $3_{03}$ & F$_2$ & (000) & $\tilde{X}^+$ & 15.98245 & 0.331877e$+$0 & 0.106230e$-$3 \\
	$1_{01}$ & F$_1$ & (010) & $\tilde{X}^+$ & & $2_{11}$ & F$_1$ & (010) & $\tilde{X}^+$ & 16.64143 & 0.516600e$+$0 & 0.186667e$-$3 \\
	$5_{14}$ & F$_2$ & (010) & $\tilde{X}^+$ & & $4_{04}$ & F$_2$ & (010) & $\tilde{X}^+$ & 16.67777 & 0.770687e$+$0 & 0.112123e$-$3 \\
	$2_{11}$ & F$_1$ & (000) & $\tilde{X}^+$ & & $3_{03}$ & F$_1$ & (000) & $\tilde{X}^+$ & 17.31330 & 0.474594e$+$0 & 0.128740e$-$3 \\
	$2_{11}$ & F$_1$ & (000) & $\tilde{X}^+$ & & $3_{03}$ & F$_2$ & (000) & $\tilde{X}^+$ & 17.31360 & 0.237352e$-$1 & 0.643882e$-$5 \\
	$5_{14}$ & F$_1$ & (010) & $\tilde{X}^+$ & & $4_{04}$ & F$_1$ & (010) & $\tilde{X}^+$ & 19.25375 & 0.948850e$+$0 & 0.176996e$-$3 \\
	$1_{01}$ & F$_2$ & (010) & $\tilde{X}^+$ & & $2_{12}$ & F$_2$ & (010) & $\tilde{X}^+$ & 22.48053 & 0.284793e$+$0 & 0.507365e$-$3 \\
	$1_{01}$ & F$_1$ & (010) & $\tilde{X}^+$ & & $2_{12}$ & F$_2$ & (010) & $\tilde{X}^+$ & 22.48056 & 0.569585e$-$1 & 0.507366e$-$4 \\
	$0_{00}$ & F$_1$ & (010) & $\tilde{X}^+$ & & $1_{10}$ & F$_1$ & (010) & $\tilde{X}^+$ & 28.99229 & 0.383520e$+$0 & 0.146557e$-$2 \\
	$1_{10}$ & F$_2$ & (000) & $\tilde{X}^+$ & & $2_{02}$ & F$_2$ & (000) & $\tilde{X}^+$ & 30.63176 & 0.136422e$+$0 & 0.614854e$-$3 \\
	$1_{11}$ & F$_1$ & (000) & $\tilde{X}^+$ & & $2_{02}$ & F$_1$ & (000) & $\tilde{X}^+$ & 33.05039 & 0.246066e$+$0 & 0.696504e$-$3 \\
	$1_{11}$ & F$_1$ & (000) & $\tilde{X}^+$ & & $2_{02}$ & F$_2$ & (000) & $\tilde{X}^+$ & 33.05048 & 0.273463e$-$1 & 0.774058e$-$4 \\
	$0_{00}$ & F$_1$ & (010) & $\tilde{X}^+$ & & $1_{10}$ & F$_2$ & (010) & $\tilde{X}^+$ & 39.53374 & 0.189105e$+$0 & 0.183223e$-$2 \\
	$1_{01}$ & F$_2$ & (010) & $\tilde{X}^+$ & & $1_{11}$ & F$_1$ & (010) & $\tilde{X}^+$ & 42.95749 & 0.958427e$-$1 & 0.119137e$-$2 \\
	$1_{01}$ & F$_1$ & (010) & $\tilde{X}^+$ & & $1_{11}$ & F$_1$ & (010) & $\tilde{X}^+$ & 42.95752 & 0.479213e$+$0 & 0.297844e$-$2 \\
	$5_{05}$ & F$_2$ & (010) & $\tilde{X}^+$ & & $5_{15}$ & F$_1$ & (010) & $\tilde{X}^+$ & 43.30037 & 0.248162e$-$1 & 0.631847e$-$4 \\
	$5_{05}$ & F$_1$ & (010) & $\tilde{X}^+$ & & $5_{15}$ & F$_1$ & (010) & $\tilde{X}^+$ & 43.30358 & 0.161381e$+$1 & 0.342486e$-$2 \\
	$4_{04}$ & F$_2$ & (010) & $\tilde{X}^+$ & & $4_{13}$ & F$_1$ & (010) & $\tilde{X}^+$ & 43.80653 & 0.308765e$-$1 & 0.101755e$-$3 \\
	$4_{04}$ & F$_1$ & (010) & $\tilde{X}^+$ & & $4_{13}$ & F$_1$ & (010) & $\tilde{X}^+$ & 43.80824 & 0.135906e$+$1 & 0.358350e$-$2 \\
	$2_{02}$ & F$_2$ & (010) & $\tilde{X}^+$ & & $2_{12}$ & F$_1$ & (010) & $\tilde{X}^+$ & 43.84737 & 0.570185e$-$1 & 0.376867e$-$3 \\
	$2_{02}$ & F$_1$ & (010) & $\tilde{X}^+$ & & $2_{12}$ & F$_1$ & (010) & $\tilde{X}^+$ & 43.84762 & 0.798375e$+$0 & 0.351800e$-$2 \\
	$3_{03}$ & F$_2$ & (010) & $\tilde{X}^+$ & & $3_{13}$ & F$_1$ & (010) & $\tilde{X}^+$ & 44.02068 & 0.402743e$-$1 & 0.179576e$-$3 \\
	$3_{03}$ & F$_1$ & (010) & $\tilde{X}^+$ & & $3_{13}$ & F$_1$ & (010) & $\tilde{X}^+$ & 44.02145 & 0.108768e$+$1 & 0.363754e$-$2 \\
	$5_{05}$ & F$_2$ & (010) & $\tilde{X}^+$ & & $5_{15}$ & F$_2$ & (010) & $\tilde{X}^+$ & 45.91836 & 0.133548e$+$1 & 0.405507e$-$2 \\
	$5_{05}$ & F$_1$ & (010) & $\tilde{X}^+$ & & $5_{15}$ & F$_2$ & (010) & $\tilde{X}^+$ & 45.92157 & 0.247426e$-$1 & 0.626205e$-$4 \\
	$4_{04}$ & F$_2$ & (010) & $\tilde{X}^+$ & & $4_{13}$ & F$_2$ & (010) & $\tilde{X}^+$ & 46.99745 & 0.107614e$+$1 & 0.437930e$-$2 \\
	$4_{04}$ & F$_1$ & (010) & $\tilde{X}^+$ & & $4_{13}$ & F$_2$ & (010) & $\tilde{X}^+$ & 46.99917 & 0.307581e$-$1 & 0.100145e$-$3 \\
	$3_{03}$ & F$_2$ & (010) & $\tilde{X}^+$ & & $3_{12}$ & F$_2$ & (010) & $\tilde{X}^+$ & 48.13497 & 0.801116e$+$0 & 0.467012e$-$2 \\
	$3_{03}$ & F$_1$ & (010) & $\tilde{X}^+$ & & $3_{12}$ & F$_2$ & (010) & $\tilde{X}^+$ & 48.13573 & 0.400660e$-$1 & 0.175183e$-$3 \\
	$2_{02}$ & F$_2$ & (010) & $\tilde{X}^+$ & & $2_{11}$ & F$_2$ & (010) & $\tilde{X}^+$ & 49.70498 & 0.509200e$+$0 & 0.490264e$-$2 \\
	$2_{02}$ & F$_1$ & (010) & $\tilde{X}^+$ & & $2_{11}$ & F$_2$ & (010) & $\tilde{X}^+$ & 49.70522 & 0.565860e$-$1 & 0.363216e$-$3 \\
	$1_{01}$ & F$_2$ & (010) & $\tilde{X}^+$ & & $1_{11}$ & F$_2$ & (010) & $\tilde{X}^+$ & 53.51002 & 0.189025e$+$0 & 0.454147e$-$2 \\
	$1_{01}$ & F$_1$ & (010) & $\tilde{X}^+$ & & $1_{11}$ & F$_2$ & (010) & $\tilde{X}^+$ & 53.51005 & 0.945124e$-$1 & 0.113537e$-$2 \\
	\hline\hline
\end{tabular}

\tablefoot{Transitions are assigned using the
rotational quantum numbers $N_{K_a K_c}$, the electron spin-rotation
label F$_{12}$, the vibrational quantum numbers $v=(v_1v_2v_3)$,
and the electronic state label $L$ of
the upper and lower levels. The transition
frequency $\nu$ in cm$^{-1}$, the line strength $S$ in Debye$^2$,
and the Einstein coefficient $A_{12}$ in s$^{-1}$ are given.}
\end{center}
\end{table*}

\FloatBarrier
\clearpage
\section{Models comparisons in the 410-1010~K range}
%________________________________________________________
\begin{figure*}[!h]
	\begin{center}
		\includegraphics[width=0.8\columnwidth,angle=0]{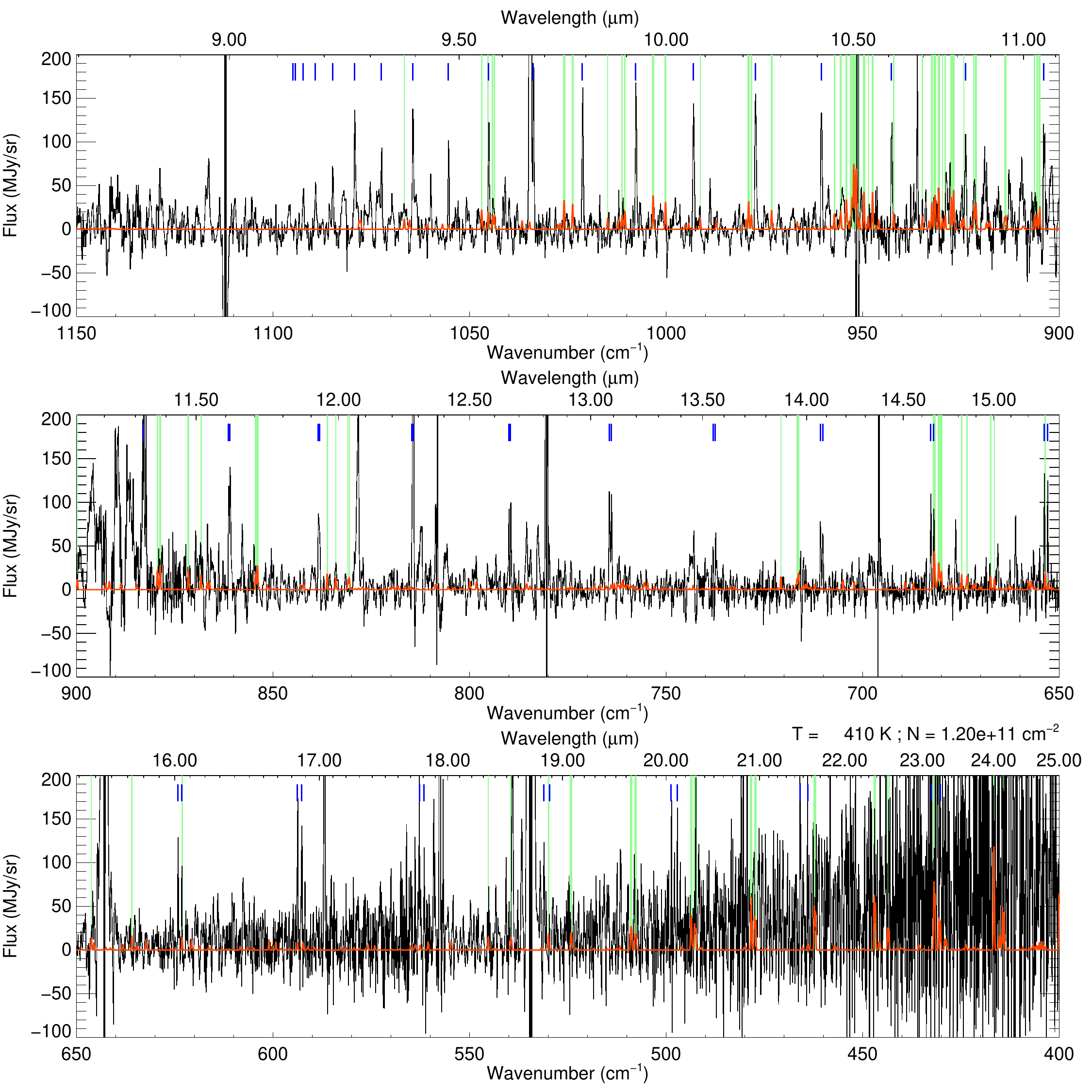}
		\caption{Comparison of the irradiated disk d203-506 continuum subtracted JWST spectrum to the CH$_2^+$ calculated LTE spectrum at 410~K with a column density limit corresponding to $1.2\times10^{11}$~cm$^{-2}$ (red). The series of emission lines labeled with blue vertical marks corresponds to OH emission already reported in
			The vertical green regions indicate the CH$_2^+$ lines used 
			to constrain the upper limit derived at that excitation temperature.}
		\label{Figure_upper_limit_ch2_plus_410K}
	\end{center}
\end{figure*}
%________________________________________________________

%________________________________________________________
\begin{figure*}[!h]
\begin{center}
\includegraphics[width=0.8\columnwidth,angle=0]{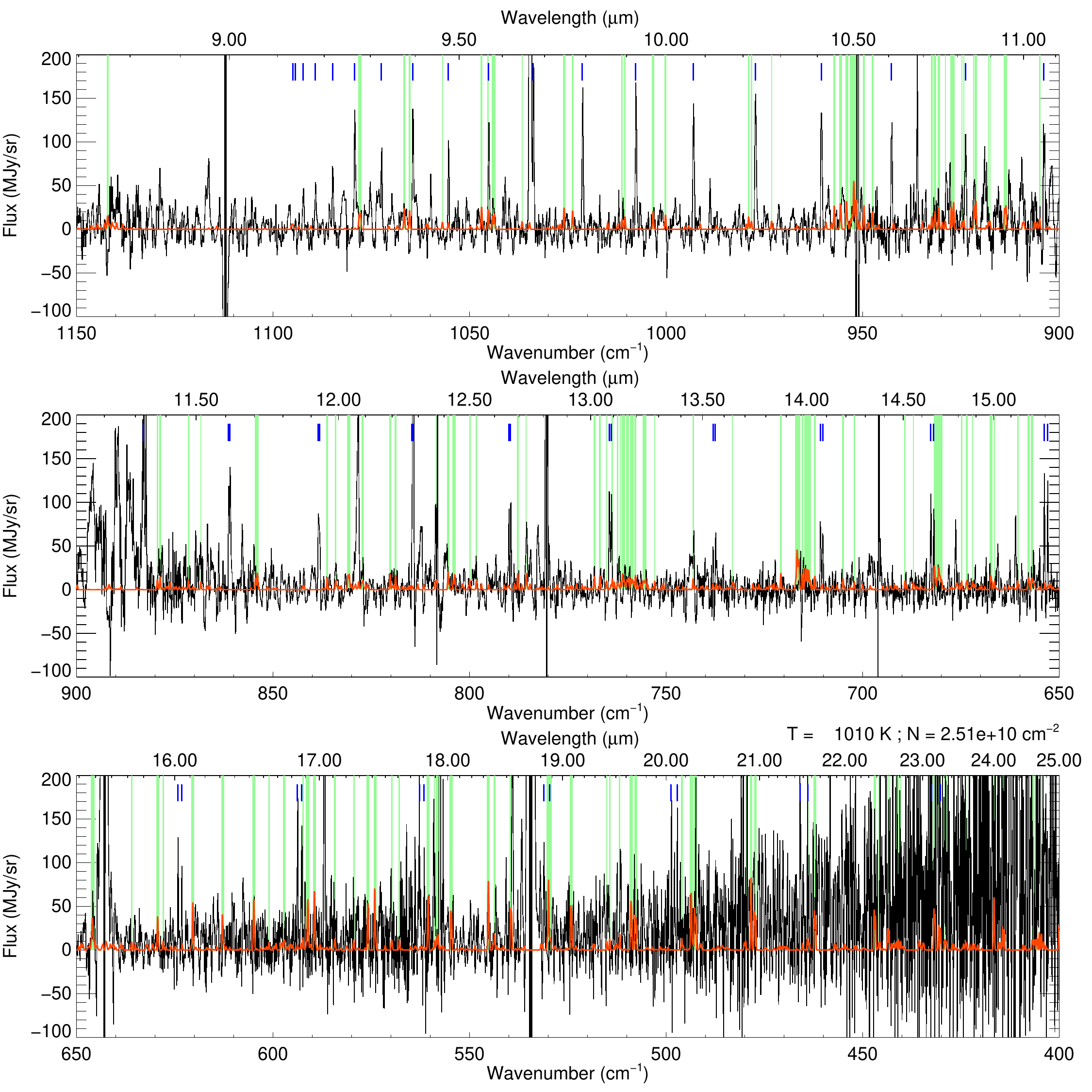}
\caption{Comparison of the irradiated disk d203-506 continuum subtracted JWST spectrum to the CH$_2^+$ calculated LTE spectrum at 1010~K with a column density limit corresponding to $2.5\times10^{10}$~cm$^{-2}$ (red). The series of emission lines labeled with blue vertical marks corresponds to OH emission already reported in \cite{Zannese_2024}. 
The vertical green regions indicate the CH$_2^+$ lines used 
to constrain on the upper limit derived at that excitation temperature.}
\label{Figure_upper_limit_ch2_plus_1010K}
\end{center}
\end{figure*}

\end{document}